\documentstyle[11pt,newpasp,twoside,epsf]{article}
\def\edcomment#1{\iffalse\marginpar{\raggedright\sl#1\/}\else\relax\fi}
\reversemarginpar

\begin{document}
\title{Brightest Cluster Galaxy Formation in Low-Redshift Galaxy Clusters}
\author{Wayne A. Barkhouse and H.K.C. Yee}
\affil{Department of Astronomy \& Astrophysics, University of Toronto, 60 
St. George Street, Toronto, ON, M5S 1A7, Canada}
\author{O. L\'{o}pez-Cruz}
\affil{INAOE-Tonantzintla, Tonantzintla, Pue., Mexico}

\begin{abstract}
We present the composite luminosity function of a sample of 17 low-redshift 
galaxy clusters. Our luminosity functions have been measured for an inner 
region ($r \leq r_{200}$) and an outer region ($r_{200} \leq r \leq 
2r_{200}$) centred on each Brightest Cluster Galaxy. 
The inner region luminosity function has a significantly 
flatter faint-end slope ($\alpha_{inner}= -1.81 \pm 0.02$) than the outer 
region ($\alpha_{outer}= -2.07 \pm 0.02$). These results are consistent with 
the hypothesis that a large fraction of dwarf galaxies near cluster centres 
are being tidally disrupted.  
\end{abstract}

\section{Introduction}

Galaxy clusters provide a laboratory in which one can study the effects of 
galaxy formation and evolution over time. A detailed understanding of 
low-redshift clusters would furnish us with a benchmark in which to compare 
observations of higher redshift galaxy clusters. Cluster properties, such as 
luminosity functions, have the potential to yield valuable clues concerning 
the dynamical evolution of cluster galaxies, including the formation of 
Brightest Cluster Galaxies (BCGs). 

Recent studies have concentrated on the merging process as an efficient way 
of creating BCGs (Merritt 1984; Schombert 1988). The formation of cD-like 
galaxies (we define cD galaxies as those with an extended halo
(e.g., Schombert 1988) as well as those galaxies which are halo-dominated; 
Brown 1997) has been problematic since the merging process cannot account 
for the presence of an extended halo (Merritt 1984; Tremaine 1990). 

From a survey of 45 low-redshift ($0.04 \leq z \leq 0.18$) galaxy clusters, 
L\'{o}pez-Cruz et al. 1997 proposed that the envelope of cD galaxies was 
created through the process of tidal disruption of dwarf galaxies (e.g., 
dwarf spheroidals). Evidence for this hypothesis was obtained by 
examining cluster luminosity functions (LFs). L\'{o}pez-Cruz et al. (1997) 
found that clusters which contain a cD galaxy (Bautz Morgan types I, I-II; 
Rood-Sastry type ``cd'') had flatter faint-end slopes (as parameterized by the 
Schechter function; Schechter 1976) than those which did not. Further support 
for this model was found by examining the dwarf-to-giant galaxy ratio (D/G),
which was found to increase with increasing Bautz Morgan type 
(L\'{o}pez-Cruz 1997).

\section{Cluster Survey}

In order to test the dwarf galaxy disruption model, we have conducted an 
imaging survey of low-redshift ($0.02 \leq z \leq 0.04$) Abell clusters using 
the KPNO 0.9m telescope + 8k mosaic camera. We have obtain R band images 
for 27 clusters, of which a sub-sample of 11 were also observed in the B band. 
The 8k mosaic camera has a field-of-view of approximately one square 
degree and allows us to cover a region 2 to 4 Mpc in length, centred on 
each cluster.

The purpose of this survey is to test the dwarf galaxy disruption hypothesis 
by; 1) sampling cluster luminosity functions to a depth adequate to measure 
the faint-end slope, 2) measuring cluster LFs over a wide area in order to 
sample the LF as a function of cluster-centric radius, 3) determining the 
colour gradient of the BCG envelope and compare it to the radial colour 
gradient 
of the dwarf galaxy population, and 4) measuring the properties 
(LFs, spatial distributions, etc.) of morphologically selected galaxy 
sub-populations.

In this paper we concentrate on results based on the analysis of 
cluster luminosity functions (we use $H_{o}= 50~km~s^{-1}~Mpc^{-1}$ and 
$q_{o}=0$ throughout). 
\subsection{Data Reductions and Photometry}
  
Object detection, classification, and photometry was performed using PPP
(Yee 1991). All bright, early-type galaxies were modeled and subtracted 
from their parent image using the same procedure as in Brown (1997). 
This allowed us to accurately measure 
the magnitudes of overlapping fainter galaxies which becomes a serious 
problem in the cluster core. Complete photometric catalogs of each cluster 
contain approximately 25,000 objects. 

Since we do not have redshifts for every cluster galaxy, luminosity 
functions are created from the statistical subtraction of background 
galaxies. The background correction was determined by observing six controls 
fields, covering a total of six square degrees in area.  

\section{Luminosity Functions}

The direct comparison of LFs for various surveys has been problematic. Many 
studies have relied on photographic plates and have compared cluster 
LFs measured using a different radial cutoff from the BCG (e.g., Lugger 1986).
Since we expect that environment may have an influence on the overall 
galaxy population 
mixture (morphology-density relation; Dressler 1980), it is prudent to 
compare cluster LFs using the same dynamical radius. For this study, we have 
chosen to measure our LFs using the $r_{200}$ radius (the radius within which 
the average density is 200 times the critical density). 
Since we lack radial velocity measurements, we use the 
correlation between $r_{200}$ and Bgc (a richness indicator; 
Yee \& L\'{o}pez-Cruz 1999) as measured for the CNOC 1 cluster survey 
(Carlberg, Yee, \& Ellingson 1997). 
We find that as cluster richness increases, the value of $r_{200}$ also 
increases. Using this relationship, we can estimate a value of 
$r_{200}$ for a given value of Bgc (which we can calculate from our 
photometry). In this way, we are able to 
make a more direct and robust comparison of cluster LFs.

In an effort to improve the signal-to-noise of cluster galaxies over 
fore/background galaxies, we impose a colour criteria from which we select 
cluster members. Cluster colour-magnitude diagrams 
clearly show the well known colour-magnitude relation (CMR; Baum 1959) for 
early-type cluster galaxies. The colour-magnitude diagram can reveal the 
presence of background contaminating clusters (L\'{o}pez-Cruz 
1997). We have used the colour-magnitude diagram to establish an upper 
envelope in which to reject redder background galaxies. 
 
\subsection{Composite Luminosity Function}

We computed the composite LF for two radial bins centred on the BCG. The 
inner area was chosen to have a radius of  $r_{200}$, and the outer 
bin had an inner radius of $r_{200}$ and an outer radius  $2r_{200}$. 
Clusters whose photometry was 100\% complete to $M_{R}=-16.0$, and which 
had complete imaging out to a radius of $2r_{200}$, were included in the 
final sample. This sample consisted of 17 clusters, 7 of which have colour 
information which allowed us to impose a colour selection criteria based 
on the cluster's CMR. All magnitudes were K corrected based on 
Coleman, Wu, \& Weedman (1980) and corrected for galactic extinction 
(Burstein \& Heilis 1982).
 
Individual cluster LFs were constructed through the statistical subtraction 
of background galaxy counts as measured from six control fields. The rejection 
criteria used for each cluster was also applied to the background counts 
(including normalization with respect to cluster area). Each cluster LF was 
then summed to produce a composite LF for each radial bin. 

Figure 1 shows the composite LF computed for the inner and outer radial 
regions. It is clear from this figure that a single Schechter function is 
inadequate to describe the composite LFs. We have therefore used the sum of 
two Schechter functions to fit the LFs in both cases. For the inner radial bin 
($r \leq r_{200}$), we find a faint-end slope of 
$\alpha_{inner}= -1.81\pm 0.02$. The faint-end slope for the outer radial bin 
($r_{200}\leq r \leq 2r_{200}$) was measured to be 
$\alpha_{outer}=-2.07\pm 0.02$. It is clear from this result that the 
inner region has a significantly flatter faint-end slope than the outer region.

\section{Discussion and Conclusions}

Schechter (1976) described an empirical function which was shown to provide 
an adequate fit to a sample of 14 cluster LFs. Many studies have 
investigated the possibility that the LF is ``universal'' 
(e.g., Lugger 1986). The results of several recent studies (e.g., 
L\'{o}pez-Cruz 
et al. 1997; Driver, Couch, \& Phillipps 1998), clearly demonstrate that 
cluster LFs are not universal but are instead a function of environment 
(this is really just a re-statement of the density-morphology relation).

The comparison of the composite LFs in this study support the conclusion 
that cluster LFs are not universal and are in fact dependent upon 
cluster-centric radius (ie., environment). This result is also consistent 
with the dwarf galaxy disruption model since one expects that tidal forces 
acting on dwarf galaxies would be greater near the central regions of 
clusters. A reduction in the fraction of dwarf galaxies, relative to the 
remaining galaxy cluster population, would be expected for the inner 
cluster regions.

\begin{figure}[!ht]
\plottwo{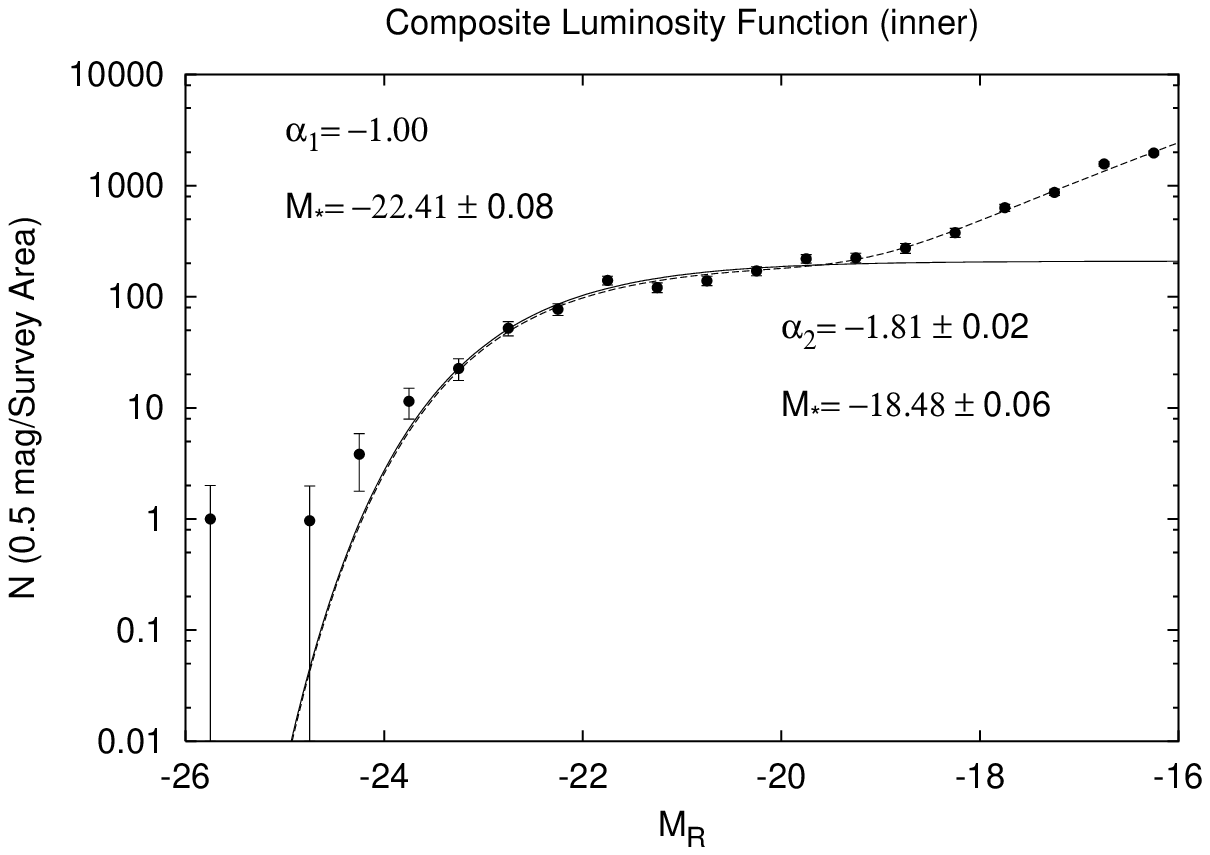}{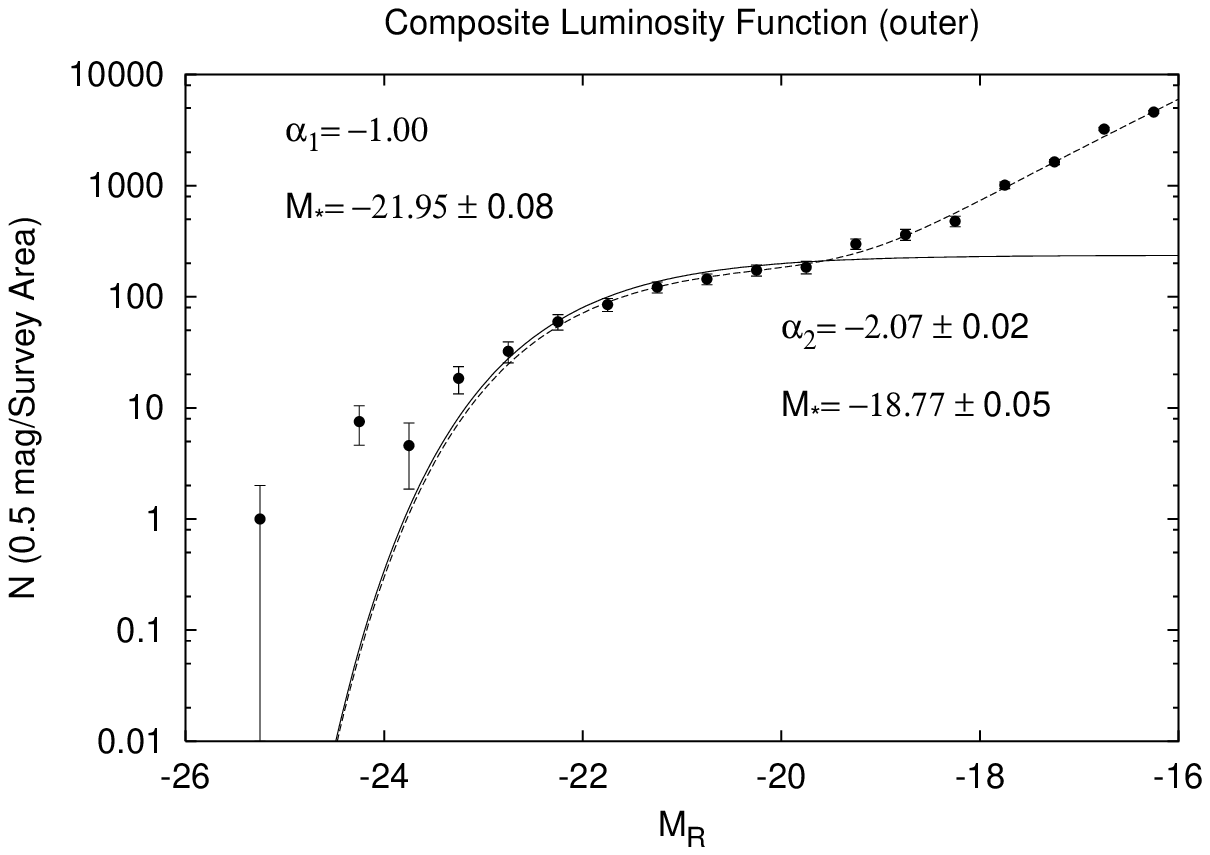}
\caption{This figure depicts the composite LF for the inner ($ r \leq 
r_{200}$) and outer ($r_{200}\leq r \leq 2r_{200}$) radial bins. The 
faint-end slope of the inner region is significantly flatter 
($\alpha_{inner}= -1.81\pm 0.02$) than the faint-end slope of the outer 
region ($\alpha_{outer}=-2.07\pm 0.02$).}
\end{figure}

\end{document}